# Superstatistical modelling of protein diffusion dynamics in bacteria

Yuichi Itto[1,2] and Christian Beck[3]

[1]ICP, Universität Stuttgart, 70569 Stuttgart, Germany

[2]Science Division, Center for General Education, Aichi Institute of Technology,

 Aichi 470-0392, Japan

[3]School of Mathematical Sciences, Queen Mary University of London,

 London E1 4NS, United Kingdom

A recent experiment [Sadoon AA, Wang Y. 2018 *Phys. Rev. E* **98**, 042411] has revealed that nucleoid associated proteins (i.e., DNA-binding proteins) exhibit highly heterogeneous diffusion processes in bacteria where not only the diffusion constant but also the anomalous diffusion exponent fluctuates for the various proteins. The distribution of displacements of such proteins is observed to take a $q$-Gaussian form, which decays as a power law. Here, a statistical model is developed for the diffusive motion of the proteins within the bacterium, based on a superstatistics with two variables. This model hierarchically takes into account the joint fluctuations of both the anomalous diffusion exponents and the diffusion constants. A fractional Brownian motion is discussed as a possible local model. Good agreement with the experimental data is obtained.

## 1. Introduction

Diffusion phenomena in living bacteria pose challenging scientific problems (see, for example, Refs. [1-5]). There is a rapidly growing development of the experimental techniques of single-particle tracking (see Ref. [6] and references therein) which has made new interesting experimental results available. Recently, highly heterogeneous diffusion processes have been observed in various experiments (see e.g. Ref. [7] and references therein), for example in the experiment of Ref. [8] (see also Ref. [9] for a recent update), for the dynamics of histonelike nucleoid-structuring proteins in living *Escherichia coli* bacteria. Such nucleoid associated proteins interact with DNA as well as with themselves and they are uniformly distributed over the bacterium. In the experiment, to measure the diffusion properties, the trajectories of these proteins, fused to fluorescent proteins, have been analyzed at individual level. An interesting observation of Ref. [8] is that the displacement, $\Delta x$, is not Gaussian distributed but obeys a $q$-Gaussian distribution [10] (sometimes also called a Pearson-type VII distribution [11])

$$P(\Delta x) \propto \left( 1 + \frac{\Delta x^2}{w^2} \right)^{-m}, \tag{1.1}$$

where $w$ is a positive quantity having the dimension of space and $m$ is a positive exponent. This density asymptotically decays with a power law, $P(\Delta x) \sim |\Delta x|^{-2m}$. This behavior is *a priori* unexpected, since it implies large probabilities of large displacements (but note that the distribution still has a finite second moment since $m \cong 2.4$, unlike the case of Lévy flights [12]). Accordingly, the behavior is in marked contrast to the results obtained in other experimental studies, for example, in Refs. [13-15], where the distribution of displacements of mRNA or chromosomal locus in living bacteria was found to be of exponential form, with a characteristic displacement (see also Ref. [16] for other cell types). Thus, the result of Ref. [8] sheds new interesting aspects onto the complexity of protein diffusion in living bacteria.

The analysis of Ref. [8] has shown that there is actually a distribution of different anomalous diffusion exponents, as well as a distribution of different diffusion constants in the bacterium. In this paper we will provide a theoretical framework for analyzing these distributions.

**Experimental observations**

Let us write for the mean square displacement of a given protein

$$\overline{\Delta x^2} \sim D_\alpha \, t^\alpha,  \quad\quad\quad (1.2)$$

where $D_\alpha$ is the diffusion constant, $\alpha$ is the (anomalous) diffusion exponent, and $t$ is elapsed time. (Here and hereafter, the notations we use are slightly different from those in Ref. [8].) Normal diffusion implies $\alpha = 1$, whereas the case with $\alpha \neq 1$ ($> 0$) corresponds to anomalous diffusion [17-19].

A fundamental experimental observation of Ref. [8] is that not only the diffusion constants but also the diffusion exponents fluctuate in a wide range. The distribution of the diffusion constants is observed to asymptotically follow a power law

$$\phi(D_\alpha) \sim D_\alpha^{-\gamma-1} \tag{1.3}$$

with $\gamma \cong 0.97$, whereas the diffusion exponents obey a rather broad distribution in the range $0 \leq \alpha \leq 2$, see Figs. 1 and 2.

Regarding the distribution in Eq. (1.3), the following points should be noted: it has been obtained for numerical values of $D_\alpha$ in the sense that the dimension is neglected, since the dimension of $D_\alpha$ changes depending on the values of $\alpha$, as can be seen in Eq. (1.2). For small elapsed time, only normal diffusion is observed and, remarkably, the diffusion constant, $D$, in this case has also been found to asymptotically obey the distribution in Eq. (1.3), denoted as $\phi(D) \sim D^{-\mu-1}$, where the exponent, $\mu$, turns out to take the value $\mu \cong 1.9$, see Fig. 3.

It may be worth pointing out that the power-law nature in Eq. (1.3) is nontrivial, since it is apparently different from the exponential law reported, for example, in Ref. [14] (see also Refs. [20,21] for an entropic approach to this law).

The mean square displacement in an ensemble average, i.e., an average of square displacement over all of the individual trajectories, has also been obtained in the measurements of Ref. [8], where both an average diffusion constant and an average diffusion exponent are determined. Then, the bacteria have been classified into three groups based on their cell age (or, equivalently cell length). As can be seen in Figs. 6(c) and 6(d) in Ref. [8], it has been observed for such groups that, in terms of the cell age, the average diffusion constant increases significantly, whereas the average diffusion exponent is approximately constant (it *increases only slightly*).

**Two-variable superstatistical treatment**

Generally, in nonequilibrium statistical physics the diffusion constant is proportional to temperature through the Einstein relation [22], hence a theory of fluctuating diffusion constants is mathematically equivalent to a theory of fluctuating temperatures. This leads us naturally to the concept of superstatistics [23]. Superstatistics is a "statistics of statistics" with largely separated time scales: a prototype example is a Brownian motion in a fluid environment in a variety of nonequilibrium stationary states [24]. The marginal distribution of the Brownian particle is written as a superposition of the statistics describing the local Brownian motion on a short time scale (where inverse temperature is locally fixed) with respect to the statistics associated with slowly fluctuating inverse temperatures on a long time scale. The superstatistical idea has been widely used for a variety of complex systems/phenomena organized by different dynamics with such a separation of time scales (see Ref. [25] for a review and Refs. [26-32] for very recent developments). Among others, the discussions in Refs. [7,33] are suitable for biological application: in the former, a fluctuating quantity is the diffusion constant, whereas it is the diffusion exponent in the latter.

The basic idea of this paper is to develop a two variable-superstatistical formalism which is then applied to model the stochastic motion of histonelike nucleoid-structuring proteins in living *Escherichia coli* bacteria. In our approach, the time scale of variation of the fluctuations of both the diffusion exponent and the inverse temperature is much larger than that of the dynamics of the protein in local areas of the bacterium, leading to associated statistics with two largely-separated time scales. In fact, the superstatistical probability densities can simply be associated with the different types of proteins that diffuse in the bacterium, thus there is naturally an ensemble of different diffusion constants and different scaling exponents of anomalous diffusion. We describe the statistical property of the protein over the bacterium as a superposition of these different statistics. For a given protein in given local areas, fractional Brownian motion [34] (a simple Gaussian stochastic process where the mean square displacement exhibits anomalous diffusion) is used as a simple local stochastic process, see also Ref. [7] and references therein. Our approach can be easily generalized to other local stochastic models, for example along the lines discussed in Ref. [35], where a Langevin equation with fluctuations of both friction and noise intensity has been studied. We show that the present theory gives rise to $q$-Gaussian (Pearson-type) distributions, in agreement with the experimental measurements. We also propose a particular form of the distribution of the anomalous diffusion exponents, again in agreement with the experimental data. Overall, our approach leads to a general characterization and effective thermodynamic description of the most important properties of complex biomolecule diffusion processes, applicable in many different contexts (see also Refs. [36-39] for further possible applications, where fluctuations of both the diffusion constant and the diffusion exponent have been experimentally observed).

## 2. Results and discussion

As mentioned earlier, for the experiments discussed here both the diffusion constant and the diffusion exponent can fluctuate in a spatio-temporal way. These fluctuations are fundamental, they do not just come from insufficient sampling of trajectories.

Generally, it is well-known that the diffusion constant for ordinary Brownian motion is proportional to temperature, the proportionality constant being the mobility of the particle. Thus, a distribution of different protein diffusion constants is formally equivalent to a distribution of different temperatures, which, by transformation of random variables, can be re-formulated as a distribution of inverse temperatures. It is well-known [24] that a $\chi^2$ distribution of inverse temperatures leads to $q$-Gaussian distributions. This we now apply to the case of protein diffusion in the bacterium with different diffusion constants. Instead of talking about a distribution of diffusion constants, we talk about the corresponding distribution of (effective) inverse temperatures, which is mathematically equivalent.

The time scale of variation of the fluctuations of both the diffusion exponent, $\alpha$, and the inverse temperature, $\beta$, is naturally much larger than that of the dynamics of the protein in a given local block (spatial region), see Appendix A. In fact different inverse temperatures can correspond to an ensemble of different proteins with different diffusion constants. Let us denote the joint distribution of both $\alpha$ and $\beta$ by $g(\alpha, \beta)$, which we can generally write as

$$g(\alpha, \beta) = g(\alpha \mid \beta) f(\beta). \tag{2.1}$$

Here, $g(\alpha \mid \beta)$ is the conditional distribution describing the probability of $\alpha$, given a value of $\beta$, and $f(\beta)$ is the marginal distribution describing the probability of $\beta$, i.e., $f(\beta) = \int d\alpha \, g(\alpha, \beta)$. For the protein (or ensemble of proteins) in a given local block, we denote the probability of finding the protein in the interval $[x, x + dx]$ at time $t$ by $P_{\alpha, \beta}(x, t) \, dx$. The protein moves from one block to another on a long time scale. Therefore, for the entire bacterium let us introduce a superstatistical ensemble of all proteins by defining the following integrated probability distribution:

$$P(x, t) = \int d\alpha \int d\beta \, g(\alpha, \beta) \, P_{\alpha, \beta}(x, t). \tag{2.2}$$

Equation (2.2) is in conformity with the viewpoint of a superstatistics with two variables: the probability distribution is expressed as a superposition of $P_{\alpha, \beta}(x, t)$ with respect to the distribution $g(\alpha, \beta)$. Thus, a statistical treatment of the fluctuations of both $\alpha$ and $\beta$ is hierarchically introduced in this way. Equation (2.2) describes a kind of superstatistical partition function, characteristic for each bacterium.

Before proceeding, we mention the following: the correlation between $\alpha$ and $\beta$ is supposed to exist at a general statistical level. Accordingly, one may wonder if the correlation is connected with a kind of fluctuation-dissipation-like relation, which is meant in the sense that $D_\alpha$ is proportional to temperature and is inversely proportional to the friction constant depending on $\alpha$. Such a relation is motivated by a similar approach described in Ref. [40]. If such a relation holds, then one can obtain a superposition of joint fluctuations in which correlations are taken into account. In the case where an ensemble of different proteins of different shapes is considered, the correlation may be non-negligible. But as will be seen in later sections, for the data studied here the correlation turns out to be weak.

Superstatistical techniques are an approximation, and one has to be clear about what types of approximations are done, and how well these are experimentally justified. In Ref. [41] a large separation of two time scales was explicitly taken into account by the use of conditional probabilities, in which the integration over the fast variable (i.e., an effective energy concerned with a local region) was taken after that of the slow variable (i.e., the inverse temperature). In Ref. [42], it was then pointed out that this procedure is opposite to the one in the adiabatic scheme (see, e.g., Ref. [43] for a relevant discussion, where a dynamical equation for a slowly fluctuating quantity was studied). In this respect, the present procedure of the integration associated with the joint fluctuations in Eq. (2.2) is seen to be consistent with the adiabatic scheme.

So far Eq. (2.2) is formal, without having determined the distributions appearing there. In the following, we discuss distributions that are consistent with the experimental data. Since $D$ is distributed according to $\phi(D)$ as noted earlier, $f(\beta)$ is found to be a power-law distribution given by $f(\beta) = |D'|\phi(D) \sim \beta^{\mu-1}$, where the relation $D \propto 1/\beta$ is assumed and the prime denotes, throughout the present work, differentiation with respect to $\beta$. In contrast to $f(\beta)$, the explicit form of $g(\alpha \,|\, \beta)$ is unspecified. We will propose in the next subsection an example of such a form, allowing us to evaluate the marginal distribution describing the fluctuations of $\alpha$, i.e., $g(\alpha) = \int d\beta\, g(\alpha, \beta)$, which can be compared with distributions observed in the experiment [see Eq. (2.12) below and Appendix D].

In the following we use the fact that the above experimentally observed power-law distribution in bacteria is only describing the behavior for small values of the inverse temperature $\beta$ (i.e. large values of temperature $T$). We are free to assume suitable behavior for larger values of $\beta$ to get agreement with $q$-Gaussians of the displacement distributions, by integrating over all $\beta$. This is the standard formalism of superstatistics for one variable as described in Ref. [24], assuming a $\chi^2$-distributed $\beta$.

Therefore, let us suppose that $f(\beta)$ is given by the following $\chi^2$ distribution

$$f(\beta) = \frac{1}{\Gamma(\mu)}\left(\frac{\mu}{\beta_0}\right)^{\mu} \beta^{\mu-1} \exp\left(-\frac{\mu\beta}{\beta_0}\right) \tag{2.3}$$

in the whole range of $\beta \in (0, \infty)$, where $\beta_0$ is the average of $\beta$ and $\Gamma(\mu)$ is the Euler gamma function. The above functional form exhibits power-law behavior for small $\beta$, as required, but also provides a cut-off for large $\beta$. We also mention that the experimental data is seen to be consistent with Eq. (2.3), see Fig. 3 as well as Appendix B.

Regarding $g(\alpha \,|\, \beta)$, let us write it as $g(\alpha \,|\, \beta) = e^{h(\alpha|\beta)}$, where $h(\alpha \,|\, \beta)$ is a suitable function. As mentioned earlier, the average anomalous diffusion exponent increases only slightly with respect to the cell age, in contrast to the average diffusion constant. This suggests the existence of a *weak* correlation between $\alpha$ and $\beta$ at the statistical level, i.e. the two variables are not fully statistically independent. Accordingly, $g(\alpha \,|\, \beta)$ has a weak dependence on $\beta$ in the sense that $h(\alpha \,|\, \beta)$ is approximately constant in the whole range of $\beta$ and accordingly its first derivative with respect to $\beta$ is small. So, we expand $h(\alpha \,|\, \beta)$ around at $\beta = \beta_0$ up to the first order of $\beta - \beta_0$:

$h(\alpha \,|\, \beta) \cong h_0(\alpha) + h_1(\alpha)(\beta - \beta_0)$, where $h_0(\alpha) \equiv h(\alpha \,|\, \beta_0)$ and $h_1(\alpha) \equiv h'(\alpha \,|\, \beta_0)$.

Thus, with $g(\alpha \,|\, \beta_0) = \exp(h_0(\alpha))$, we have the following conditional distribution:

$$g(\alpha \,|\, \beta) \sim g(\alpha \,|\, \beta_0) \exp[\,(\beta - \beta_0)\, h_1(\alpha)\,], \qquad (2.4)$$

and the weak correlation is described in this way. The weakness of dependence is guaranteed if $h_1(\alpha)$ is assumed to be small.

From Eqs. (2.3) and (2.4), the marginal distribution $g(\alpha)$ is immediately calculated to be

$$g(\alpha) \sim g(\alpha \mid \beta_0) \frac{\exp(-\beta_0 h_1(\alpha))}{[1 - (\beta_0/\mu)h_1(\alpha)]^\mu}, \qquad (2.5)$$

where the quantity $1 - (\beta_0/\mu)h_1(\alpha)$ has been assumed to be positive [which is indeed confirmed in the case of Eq. (2.12) below, see Appendix D].

**A possible model**

For $P_{\alpha,\beta}(x,t)$, as supported e.g. by the data in Ref. [8] and examined for the present case in Appendix C, we apply the approach of fractional Brownian motion [34] as a possible stochastic process. By this we mean that the probability density is given as follows [19]:

$$P_{\alpha,\beta}(x,t) = \frac{1}{\sqrt{4\pi D_{\alpha,\beta} t^\alpha}} \exp\left(-\frac{x^2}{4D_{\alpha,\beta} t^\alpha}\right), \qquad (2.6)$$

where $D_{\alpha,\beta}$ denotes the diffusion constant. As discussed in Appendix A, it is given by the relation $D_{\alpha,\beta} \sim c/(s^{\alpha}\beta)$ with $s$ and $c$ being a characteristic time and a positive constant, respectively, and it corresponds to $D_{\alpha}$ in Eq. (1.2). The effective temperature dependence of $D_{\alpha,\beta}$ can be non-trivial in experiments. Therefore, it is of interest to experimentally examine how this dependence is realized.

Substituting Eqs. (2.1), (2.3), (2.4), and (2.6) into Eq. (2.2), we obtain the following distribution:

$$P(x,t) \sim \left\langle P_{\alpha}(x,t) \exp(-\beta_0 h_1(\alpha)) \right\rangle_{\alpha}, \qquad (2.7)$$

where the symbol $\langle \bullet \rangle_{\alpha}$ denotes the average with respect to the conditional distribution $g(\alpha \mid \beta_0)$, i.e., $\langle Q \rangle_{\alpha} \equiv \int_0^2 d\alpha\, g(\alpha \mid \beta_0) Q$ and $P_{\alpha}(x,t)$ is defined by

$$P_{\alpha}(x,t) = \frac{\Gamma(\mu+1/2)}{\Gamma(\mu)} \sqrt{\frac{\beta_0}{4\pi\mu c (t/s)^{\alpha}}} \left[ 1 - \frac{\beta_0}{\mu} h_1(\alpha) + \frac{\beta_0}{4\mu c} \frac{x^2}{(t/s)^{\alpha}} \right]^{-(\mu+1/2)}, \quad (2.8)$$

provided that, following the formalism in Ref. [24], the integration over $\beta$ has been performed.

We here discuss some further experimental features of the data in Ref. [8]. Based on the experimental data presented in Figs. 6(a), 6(c), and 6(d) in Ref. [8] and Eq. (1.2), we estimate the typical spatial scale of the local block as $\sqrt{\overline{\Delta x^2}}$ at the elapsed time of $0.2\,\mathrm{s}$, which gives the first five data points (of the total of ten data points) in the mean square displacement. $\sqrt{\overline{\Delta x^2}}$ is found to be of the order of $100\,\mathrm{nm}$ for the three age groups mentioned earlier. Then, as can be seen in Fig. 4(a) in Ref. [8], the displacements of $|\Delta x| < 500\,\mathrm{nm}$ have been examined via Eq. (1.1) for a larger time, which is supposed to approximately be $0.41\,\mathrm{s}$ in Fig. 2(b) of Ref. [8], and the cell size of bacteria studied in Ref. [8] is in the range between $1\mu\mathrm{m}$ and $6\mu\mathrm{m}$. Therefore, it seems natural to consider, for such a large time, that the protein diffuses over a region of a few local blocks in the bacterium. This implies that the fluctuations of $\beta$, rather than those of $\alpha$, dominantly contribute to the protein dynamics due to its power-law nature. In the latter, we employ the average value, $\hat{\alpha}$, of the diffusion exponent for the bacteria in each group based on the experimental data in Fig. 6(d) in Ref. [8].

Based on these observations, in the region of the local blocks with constant diffusion exponent, we can take the conditional distribution in Eq. (2.4) in good approximation as follows: $g(\alpha \mid \beta) = \delta(\alpha - \hat{\alpha})$, for which $h_1(\alpha) = 0$. From this, as well as Eqs. (2.7) and (2.8), we immediately derive the following distribution:

$$P(x,t) \sim \frac{\Gamma(\mu+1/2)}{\Gamma(\mu)} \sqrt{\frac{\beta_0}{4\pi\mu c \, (t/s)^{\hat{\alpha}}}} \left[1 + \frac{\beta_0}{4\mu c} \frac{x^2}{(t/s)^{\hat{\alpha}}}\right]^{-(\mu+1/2)}. \qquad (2.9)$$

Clearly, Eq. (2.9) has the form of a $q$-Gaussian distribution. In Fig. 4, we present the plot of $P(x,t)$ in Eq. (2.9). The experimental data are nicely described by the $q$-Gaussian distribution in Eq. (2.9). Upon renaming, $\Delta x \leftrightarrow x,$ in Eq. (1.1), we therefore have the following relations:

$$m = \mu + \frac{1}{2}, \qquad (2.10)$$

$$w \propto t^{\hat{\alpha}/2}. \qquad (2.11)$$

Equation (2.10) tells us how the power-law exponent, the value of which is seen to be about $m \cong 2.4$ (or, $q \cong 1.4$ in terms of the exponent $q$ in $q$-Gaussians), originates from the diffusion-constant fluctuations, whereas Eq. (2.11) determines the time evolution of $w$.

Let us briefly pause here to mention the experimental importance of fractional Brownian motion models. Golding and Cox [1] experimentally studied the diffusion of mRNA molecules inside *E. coli* bacteria, and it seems that they already at that early stage proposed the superposition of fractional Brownian motion processes as a suitable model, which was compared with measurements. In Ref. [44], a suitable distribution of the diffusivity was employed, and anomalous diffusion with ergodicity breaking [19] was found to emerge. The work in Ref. [45] has shown that Weibull-type distributions of the diffusivity give rise to mean square displacements similar to those experimentally observed for mRNA molecules. Whereas Ref. [44] mainly clarified the mathematical setting, Ref. [45] dealt with the experimental fine-tuning.

So far we have focused our attention on the protein diffusion process in a small region of a few local blocks/areas with given local diffusion exponent. On a sufficiently long time scale, the protein will diffuse over the entire cytoplasm of the bacterium, and it is reasonable to consider that the contributions from the fluctuations of $\alpha$ are important as well, in addition to the fluctuations of $\beta$. This describes —in a superstatistical way— additional fluctuations in the anomalous diffusion exponent. The idea is that in some regions, e.g. in very crowded areas with obstacles, subdiffusion is dominant, whereas in other regions nearly ballistic motion (superdiffusion) may be possible. These possibilities fluctuate spatially over the bacterium, and this also depends on the complexity and shape of the particular protein chosen, i.e. the ensemble of all biomolecules. (For a possible relevance to the latter, see a recent work in Ref. [46], where fluctuations of the size of a polymer have been discussed in connection with fluctuating diffusivity.)

To examine Eq. (2.7) in this situation, all kinds of conditional distributions can be considered. In the following we wish to discuss a particular example form of the conditional distribution $g(\alpha \,|\, \beta)$. As discussed in Appendix D, this is given by

$$g(\alpha \mid \beta) = N(\beta) \; s^{\gamma \alpha} \exp\!\left(-\frac{\gamma \, a(\beta)}{c} s^{\alpha}\right), \tag{2.12}$$

where $a(\beta)$ is a positive quantity depending weakly on $\beta$ and $N(\beta)$ is the normalization factor. Only dimensionless numerical values of all quantities appearing are treated, since Eq. (2.12) is obtained from the distribution of dimensionless numerical values of $D_{\alpha}$. $h_1(\alpha)$ is calculated in this case to be

$$h_1(\alpha) = \frac{\gamma \, a_1}{c} (\left\langle \, s^{\alpha} \, \right\rangle_{\alpha} - s^{\alpha}), \tag{2.13}$$

where $a_1$ ($< 0$) is a small constant, representing a weak correlation.

In Fig. 2, we show the plot of the marginal distribution $g(\alpha)$ as given in Eq. (2.5) together with Eqs. (2.12) and (2.13). There, it is observed that this distribution fits the experimental data quite well: in particular, it takes on a maximum value near $\alpha = 0.6$ and a local minimum value near $\alpha = 0.1$, respectively.

We think that for the data we use there is evidence for sufficient sampling of the trajectories. According to Ref. [8], trajectories with a minimum length of 10 frames (from the total of 20000 frames in the resulting movies) were analyzed for calculating the mean square displacements, some of which are based on trajectories that are quite long, see Fig.

3(a) in Ref. [8]. Fluctuations may sometimes just originate from a lack of precision and insufficient statistics, as discussed in Refs. [47-49], where estimation techniques for this have been developed. We believe that in our case here we have more fundamental fluctuations, not caused by insufficient sampling. The function $g(\alpha)$ in Eq. (2.5) fits quite well with the experimental data as mentioned above, under the assumption (see Appendix D) that given a value of $\beta$, the distribution of $D_\alpha$ takes the form in Eq. (B2) leading to Eq. (1.3). There is robust and reproducible behavior of the diffusion-constant fluctuations caused by medium heterogeneity (similarly, also diffusion-exponent fluctuations), with a possible additional correction from ensemble heterogeneity.

Further experimental data of protein anomalous diffusion dynamics can help to build the optimum model for a given experiment, possibly checking for universal and non-universal properties, i.e. properties that are observed for all bacteria and other properties that are very specific to a given specific experiment.

Importantly, when using superstatistical descriptions one has to check under which conditions such a description is a valid approximation for a given dynamics or experimental realization. In Ref. [50] (see also, e.g., Refs. [7,51]), a superstatistical description based on a gamma distribution of diffusivity was used for modelling the non-Gaussianity of displacements. For the particular dynamical model studied in Ref. [50] it was shown that such a description is appropriate only for short time scales, on which the diffusivity is considered not to be changing too much. Now, in the present work we see that superstatistical modelling seems to make sense also for longer time scales, fitting experimental data well, in fact for two simultaneous observables such as the diffusion constant and the diffusion exponent, although one still needs to investigate in more detail what the microscopic origin of the observed 2-variable superstatistical behavior in bacteria is. Further studies along these lines, and further comparison with further sets of experimental data, are needed to ultimately answer this question.

## 3. Conclusion

We have developed an effective superstatistical kinetic theory for describing the diffusion dynamics of an ensemble of complex biomolecules, in our case applied to the specific example of histonelike nucleoid-structuring proteins in living *Escherichia coli* bacteria. This model hierarchically takes into account both the diffusion-exponent fluctuations and the temperature (diffusion constant) fluctuations. We have shown that the theory naturally contains the $q$-Gaussian (Pearson-type) distributions often observed in experiments, for which the temperature fluctuations play a crucial role, representing local changes of the diffusion constant. The approach of fractional Brownian motion has been applied as a local stochastic process, representing the presence of an additional spectrum of anomalous diffusion exponents, which is important to consider in the most general and most complex cases describing the full contents of biomolecules in the bacterium. Proposing a concrete statistical form of the diffusion-exponent fluctuations, we have discussed the protein diffusion dynamics on a long time scale, for which the existence of the weak correlation between both the fluctuations is essential, described by a generalized two-parameter superstatistical formalism. We believe this formalism is quite generally applicable to a large variety of complex anomalous diffusion processes in small biological systems, and experimentally testable in future experiments.

**Authors' contributions.** Y.I. and C.B. jointly designed research and wrote the paper.

**Competing interests.** The authors declare no conflict of competing interest.

**Acknowledgement.** The present research has been designed jointly by Y.I. and C.B. and it was completed while Y.I. stayed at the Institut für Computerphysik, Universität Stuttgart. Y.I. would like to thank the Institut für Computerphysik for their warm hospitality.

# Appendix A

## Superstatistical fluctuations of the diffusion constant

Consider the stochastic motion of the protein over the bacterium, which is regarded as a complex medium for the diffusion of the proteins. This medium is then divided into many small spatial regions or "blocks", in each of which the protein exhibits heterogeneous diffusion according to Eq. (1.2), leading to variations of both $\alpha$ and $\beta$ depending on the local blocks. These superstatistical fluctuations are considered to give rise to the fluctuations of $D_\alpha$ as follows. The fact noted earlier that $D$ is distributed according to $\phi(D)$ suggests that this fluctuation comes from displacement of the protein rather than the characteristic time being required for displacement, since such a characteristic time is constant in the random walk picture for normal diffusion [52]. The situation may be the same for $D_\alpha$ due to the same power-law nature of diffusivity. The power-law nature implies that $D$ and $D_\alpha$ share a similar origin in their fluctuations. In fact it is clear that one can define diffusion constants for both normal and anomalous diffusion, just their dimension is different. The diffusion constant for normal diffusion is proportional to temperature, i.e., $D \propto 1/\beta$, as in the Einstein relation [22], such an origin is expected to be formally related to temperature. To investigate this point in more detail for anomalous diffusion, let us note the experimental fact [8] that for the mean square displacement in the ensemble average, the numerical value of the diffusion constant in the case of normal diffusion observed for small elapsed time is three times larger than that in the case of anomalous diffusion observed for large elapsed time, at least for the data set that we study here. This indicates, for a given individual trajectory, that $\Delta^2 \sim 3\Delta_\alpha^2$, where $\Delta$ ($\Delta_\alpha$) stands for displacement. [This symbol for the displacement should not be mixed with the one in Eq. (1.1)]. Therefore, denoting $D = \Delta^2/s$ and $D_\alpha = \Delta_\alpha^2/s^\alpha$ with $s$ being a positive constant describing the characteristic time mentioned above, these observations allow us to evaluate $D_\alpha \sim (D/3)s^{1-\alpha}$. Thus, $D_\alpha$ is given by

$$D_\alpha = D_{\alpha,\beta} \sim \frac{c}{s^\alpha \beta} \qquad (A1)$$

with $c$ being a positive constant.

From the above, the value of $\beta_0/c$, which turns out to be relevant through the present work, is estimated as follows. In the case of normal diffusion, the average of $D_{\alpha,\beta}$ over the distribution $f(\beta)$ in Eq. (2.3) is given by

$$\int_0^\infty d\beta \, f(\beta) D_{1,\beta} \sim \frac{\mu}{\mu-1} \frac{c}{s\beta_0}. \qquad (A2)$$

This should take on approximately the value of one third of $24 \times 10^3 \, \text{nm}^2/\text{s}$, for the example experiments we look at here, which is the average value of $D$, and $s = 0.045\,\text{s}$ [8], hence it is found that $\beta_0/c \sim 5.9 \times 10^{-3}\,\text{nm}^{-2}$. Also, the average of $D_{\alpha,\beta}$ over the joint fluctuations given in Eqs. (2.3) and (2.4) can be calculated and is given by

$$\int_0^2 d\alpha \int_0^\infty d\beta \, g(\alpha,\beta) D_{\alpha,\beta} \sim \frac{\mu}{\mu-1} \frac{c}{\beta_0} \left\langle \frac{\exp(-\beta_0 h_1(\alpha))}{s^\alpha [1-(\beta_0/\mu) h_1(\alpha)]^{\mu-1}} \right\rangle_\alpha. \qquad (A3)$$

Using Eqs. (2.12) and (2.13), the value turns out to be approximately given by $10 \times 10^3\,\text{nm}^2/\text{s}^\alpha$. This value is close to the average value of $D_\alpha$ measured in the experiment, which is $8.0 \times 10^3 \, \text{nm}^2/\text{s}^\alpha$ [8]. This observation is seen to support the relation in Eq. (A1).

**Appendix B**

**The fluctuation distribution of the diffusion constant**

The left-bottom inset in Fig. 2(d) in Ref. [8] is seen to imply that the experimental data of the diffusion constant $D$ [$\mu\text{m}^2/\text{s}$] can be fitted with a slightly curved line, since the data point at $D = 0.03 \, \mu\text{m}^2/\text{s}$ is below the red dashed line presented there, whereas the data points around at $D = 0.1 \, \mu\text{m}^2/\text{s}$ are above the red dashed line. Accordingly, we suppose that such a curved line is described by the following inverse gamma distribution

$$\phi(D) \propto A^\mu D^{-\mu-1} \exp\left(-\frac{\mu A}{D}\right) \qquad (B1)$$

in the interval $\varepsilon \leq D \leq \kappa$, where $A$ is a positive constant having the dimension of the diffusion constant, $\varepsilon$ and $\kappa$ are lower and upper bounds on $D$, respectively, see Fig. 3. Clearly, this distribution decays as a power law for large $D$.

Here, we employ the assumption that the range of $\beta$ in Eq. (2.3) is unbounded, although the values of $\varepsilon$ and $\kappa$ in practice (i.e. in true experiments) are of course small and large, respectively, but still finite. To illustrate this aspect, we examine the influence of $\kappa$ in our present theory. Let us compare the average value of $D$ in the two cases with $A = 0.011 \mu \text{m}^2 / \text{s}$ and $\varepsilon = 0.025 \mu \text{m}^2 / \text{s}$ (see Fig. 3), which is given by $0.051 \mu \text{m}^2 / \text{s}$ in the case of $\kappa = 0.205 \mu \text{m}^2 / \text{s}$, and by $0.062 \mu \text{m}^2 / \text{s}$ in the limit $\kappa \to \infty$, showing that these two values are close to each other. Since $D$ appears in the denominator in the exponential factor in Eqs. (2.4), (2.6), and (2.12), the contribution from $D$ becomes small as it increases. From this we conclude that the influence of $\kappa$ is quite negligible.

From the experimental data in Fig. 2(d) in Ref. [8], the situation seems to be similar for the dimensionless numerical values of $D_\alpha$ [ $\mu \text{m}^2 / \text{s}^\alpha$ ]. That is, it is seen that the data point at $D_\alpha = 0.01 \ \mu \text{m}^2 / \text{s}^\alpha$ is below the red dashed line presented there, whereas almost all of the data points between $D_\alpha = 0.02 \ \mu \text{m}^2 / \text{s}^\alpha$ and $D_\alpha = 0.1 \ \mu \text{m}^2 / \text{s}^\alpha$ are above the red dashed line, implying that the experimental data can be fitted by a slightly curved line. As this curved line, we shall take the inverse gamma distribution given by

$$\phi(D_\alpha) \propto \tilde{A}^\gamma D_\alpha^{-\gamma-1} \exp\left( -\frac{\gamma \tilde{A}}{D_\alpha} \right) \tag{B2}$$

in the interval $\tilde{\varepsilon} \le D_\alpha \le \tilde{\kappa}$, where $\tilde{A}$ is a dimensionless positive constant, $\tilde{\varepsilon}$ and $\tilde{\kappa}$ are lower and upper bounds of $D_\alpha$, respectively. As shown in Fig. 1, this distribution is seen to fit well, exhibiting the power-law behavior in Eq. (1.3) for large $D_\alpha$.

As will be shown in Appendix D, the distribution in Eq. (B2) turns out to play a key role for obtaining the form of the conditional distribution $g(\alpha \,|\, \beta)$.

## Appendix C

## A possible local stochastic process

Fractional Brownian motion [34,53] has been employed as the underlying process describing anomalous diffusion observed in a wide class of crowded fluid/biological systems, see Refs. [54-57], for example. This fact naturally motivates us to examine if the approach of fractional Brownian motion is appropriate as a model for the typical local diffusion dynamics in a bacterium, providing a concrete formula for $P_{\alpha,\beta}(x,t)$.

A process of fractional Brownian motion is described —based on that of the ordinary Brownian motion denoted by $B(t)$— in such a way that past increments of $B(t)$ are incorporated into the process: $B_H(t) = t_0^{-H+1/2} \left\{ {}_{-\infty}I_t^{H+1/2}[\,\xi(t)\,] - {}_{-\infty}I_0^{H+1/2}[\,\xi(t)\,] \right\}$, where $t_0$ is a positive constant having the dimension of time, $\xi(t)$ is the unbiased Gaussian white noise satisfying $\xi(t)\,dt = dB(t)$, and ${}_{-\infty}I_t^{\sigma}[\,\rho(t)\,] \equiv (1/\Gamma(\sigma)) \int_{-\infty}^{t} d\tau\,(t-\tau)^{\sigma-1}\rho(\tau)$ is the Riemann-Liouville fractional integral operator [58] with $\sigma > 0$. Here, $H$ is referred to as the Hurst exponent and it satisfies $0 < H < 1$. Normal diffusion is realized in the case of $H = 1/2$, whereas the case of $H \neq 1/2$ describes anomalous diffusion. So, if $B_H(t)$ describes the individual trajectory in a given local region, then $\alpha = 2H$ should hold in Eq. (1.2), [where it is understood that the case of $\alpha = 0$ ($\alpha = 2$) is realized in the limit $H \to 0$ ($H \to 1$)]. In our present discussion, the following three features of the process of fractional Brownian motion are relevant. Firstly, the velocity autocorrelation function in the process becomes negative (positive) if $0 < H < 1/2$ ($1/2 < H < 1$) [19]. This is consistent with the experimental result presented in Fig. 5(a) in Ref. [8], where it is clearly seen that the velocity autocorrelation function of the proteins becomes negative for small elapsed time. It should be also noticed that the case of $0 < \alpha < 1$ has been predominantly observed. Secondly, the diffusion processes are assumed to be ergodic, since the protein can diffuse almost anywhere, in the sense [19,59] that the mean square displacement obtained through the position autocorrelation function is equivalent to the one derived based on the probability distribution. Lastly, the process does not exhibit any aging phenomenon, since the sequence of the increments of $B_H(t)$ is stationary [19]. Regarding this point, we should say that for the experimental data [8] there is a dependence on cell age, but this can be incorporated in the choice of the effective temperature relevant for each cell age group.

In Ref. [60], the displacement autocorrelation function has been calculated for the process of fractional Brownian motion. For large elapsed time, the function becomes negative (positive) if $0 < \alpha < 1$ ($1 < \alpha < 2$). Therefore, the corresponding fractional approach is suitable if such a behavior is observed for the experimental data.

Taking into account all of these considerations, we see that the model of local fractional Brownian motion is a good one for the local dynamics modelling in bacteria and other small complex systems. It simply needs to be amended by the detailed properties of the diffusion-constant distribution and diffusion-exponent distribution.

**Appendix D**

**The conditional fluctuation distribution of the diffusion exponent**

Here, we present a particular form of the distribution $g(\alpha \,|\, \beta)$, which allows us to derive the form of the marginal distribution $g(\alpha)$. Given a value of $\beta$, an allowed range of dimensionless numerical values of $D_\alpha$ can be determined through the diffusion exponent $\alpha$ in Eq. (A1). Let us assume that the normalized probability distribution of such values takes on the form of the inverse gamma distribution in Eq. (B2). Accordingly, $g(\alpha \,|\, \beta)$ is given by $g(\alpha \,|\, \beta) = \left| \partial D_{\alpha,\beta} / \partial \alpha \right| \phi(D_\alpha)$, where $\alpha$ is considered to be distributed in the interval $0 < \alpha < 2$ due to the fact [8] that the average diffusion exponent is approximately constant in terms of the cell age, which implies that the interval to be taken does not drastically change. Since $g(\alpha \,|\, \beta)$ should have a weak dependence on $\beta$, we consider that $\tilde{A}$ in this case depends on $\beta$ in such a way that $\tilde{A}(\beta) = a(\beta) / \beta$, where $a(\beta) \cong a_0 + a_1(\beta - \beta_0)$ with $a_0 \equiv a(\beta_0)$ and $a_1 \equiv a'(\beta_0)$. Here, $a_1$ should be small, realizing the weakness. Thus, $g(\alpha \,|\, \beta)$ can be expressed as the distribution in Eq. (2.12) with $N(\beta)$ being the normalization factor given by

$$N(\beta) = \frac{\left( \gamma a(\beta) / c \right)^\gamma \ln s}{\Gamma(\gamma, \gamma a(\beta) / c) - \Gamma(\gamma, s^2 \gamma a(\beta) / c)}, \qquad (D1)$$

where $\Gamma(k, y)$ is the incomplete gamma function defined by $\Gamma(k, y) = \int_y^\infty du\, u^{k-1} e^{-u}$. Using this, $\langle s^\alpha \rangle_\alpha$ in Eq. (2.13) is found to be given by

$$\langle s^\alpha \rangle_\alpha = \frac{c}{\gamma a_0} \frac{\Gamma(1+\gamma,\, \gamma a_0 / c) - \Gamma(1+\gamma,\, s^2 \gamma a_0 / c)}{\Gamma(\gamma,\, \gamma a_0 / c) - \Gamma(\gamma,\, s^2 \gamma a_0 / c)}. \tag{D2}$$

$g(\alpha \mid \beta)$ in Eq. (2.12) is peaked at $\alpha = \alpha_*(\beta) = [\ln(c / a(\beta))] / \ln s$, and the center of this distribution should tend to approach the origin $\alpha = 0$ as $\beta$ increases, since the average diffusion exponent slightly increases with respect to the cell age [8]. Accordingly, the condition $\alpha_*' < 0$ requires $a_1$ to be negative.

Keeping this in mind, it is found that the marginal distribution $g(\alpha)$ in Eq. (2.5) with Eqs. (2.12) and (2.13) takes on a maximum value and a local minimum value, respectively, at $\alpha = \alpha_+$ and $\alpha = \alpha_-$ $(< \alpha_+)$:

$$\alpha_+ = \frac{1}{\ln s} \left[ \ln \left( \frac{a_0}{c} + \delta_+ - \sqrt{\left( \frac{a_0}{c} + \delta_- \right)^2 - \frac{4\gamma}{\mu} \left( \frac{a_1 \beta_0}{c} \right)^2} \right) - \ln \frac{\delta_+ + \delta_-}{\langle s^\alpha \rangle_\alpha} \right] \tag{D3}$$

and

$$\alpha_- = \frac{1}{\ln s}\left[\ln\left(\frac{a_0}{c} + \delta_+ + \sqrt{\left(\frac{a_0}{c} + \delta_-\right)^2 - \frac{4\gamma}{\mu}\left(\frac{a_1\beta_0}{c}\right)^2}\right) - \ln\frac{\delta_+ + \delta_-}{\langle s^\alpha\rangle_\alpha}\right], \qquad \text{(D4)}$$

where $\delta_+$ and $\delta_-$ $(< \delta_+)$ are defined by

$$\delta_+ = \frac{\gamma}{\mu}\frac{a_1\beta_0}{c}\left[\langle s^\alpha\rangle_\alpha\left(\frac{a_1\beta_0}{c} - \frac{a_0}{c}\right) - 1\right] \qquad \text{(D5)}$$

and

$$\delta_- = \frac{\gamma}{\mu}\frac{a_1\beta_0}{c}\left[\langle s^\alpha\rangle_\alpha\left(\frac{a_1\beta_0}{c} - \frac{a_0}{c}\right) + 1\right], \qquad \text{(D6)}$$

where the quantity inside the square root in Eqs. (D3) and (D4) is assumed to be positive (which is confirmed in the present case, see below).

In Fig. 2, $a_0$ and $a_1$ are taken in such a way that $\alpha_+ \cong 0.6$ and $\alpha_- \cong 0.05$, which is very similar to the experimental result [8], see the histogram in Fig. 2.

The experimental data imply that the quantity $1-(\beta_0/\mu)h_1(\alpha)$ with Eq. (2.13) in Eqs. (2.5), (2.8), and (A3) takes on values between $0.12$ and $1.16$, whereas the quantity inside the square root in Eqs. (D3) and (D4) has approximately the value of $38$, i.e. it is positive, as required.

# Figures

**Figure 1**

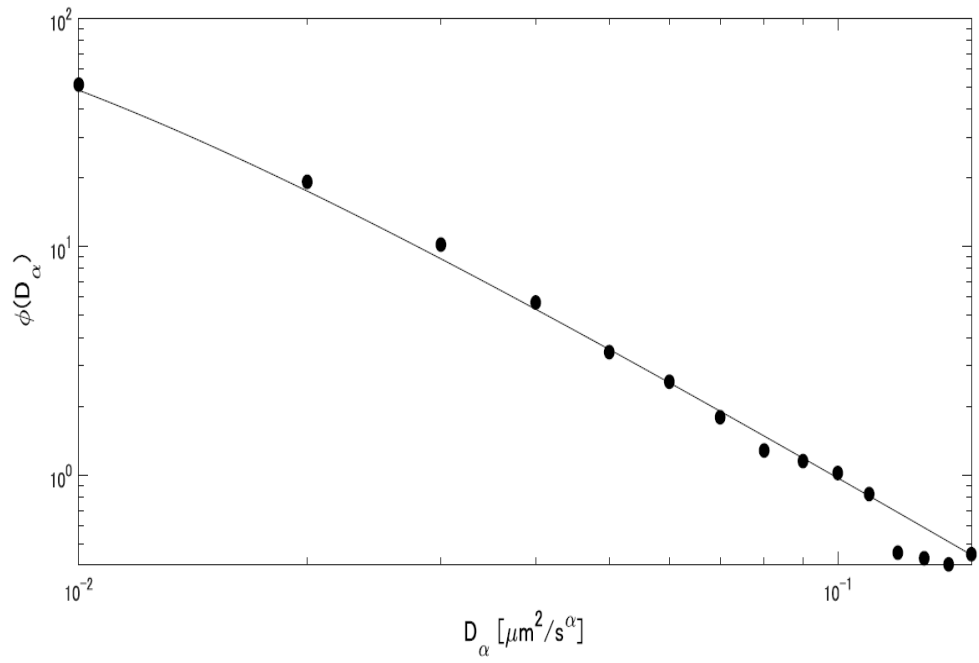

**Figure 2**

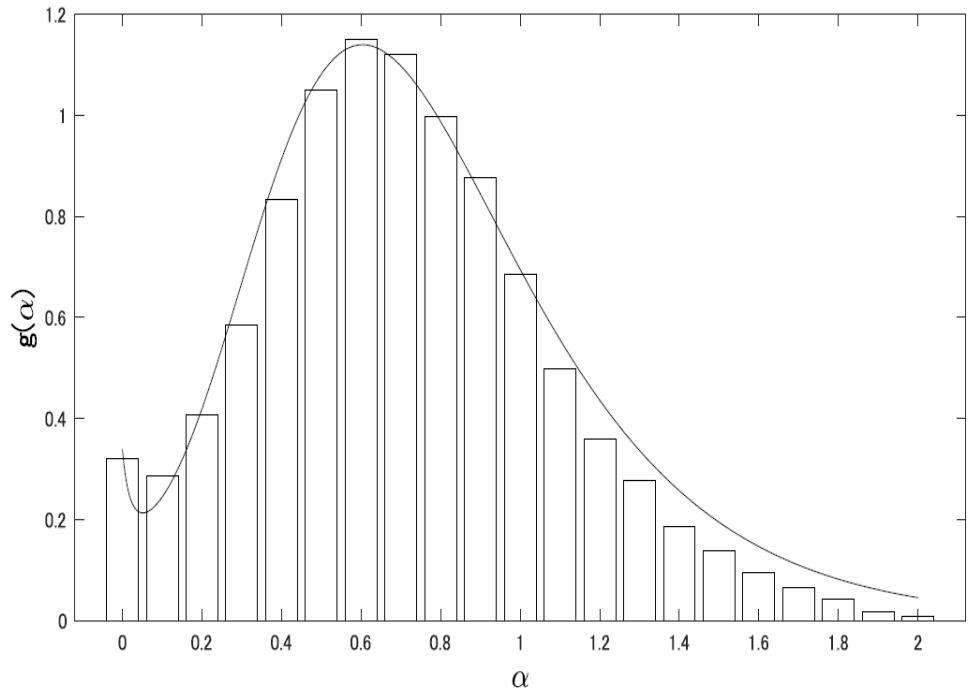

**Figure 3**

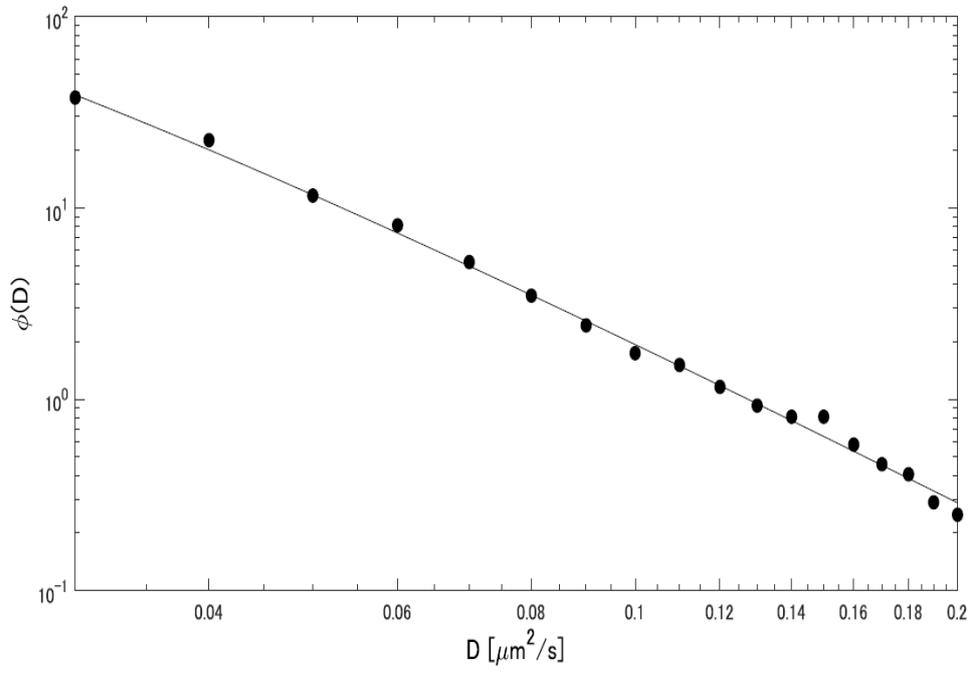

**Figure 4**

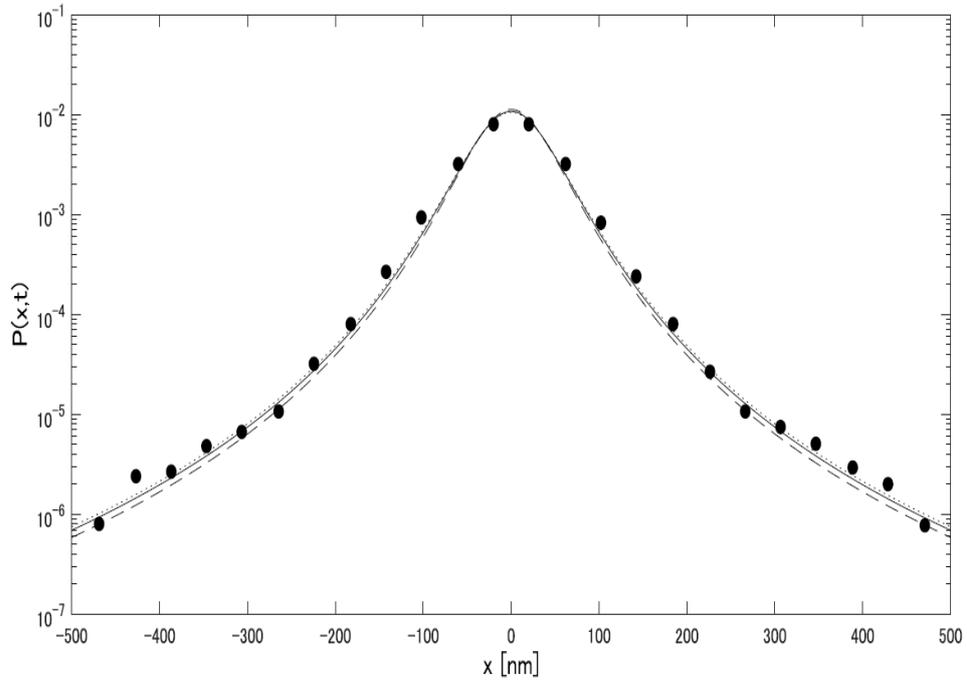

# Figure Captions

**Figure 1**

Log-log plot of the normalized distribution $\phi(D_\alpha)$ of the numerical values of $D_\alpha$ [$\mu m^2/s^\alpha$]. The histogram is based on the experimental data of Fig. 2(d) in Ref. [8]. The solid line shows the inverse gamma distribution, Eq. (B2), with $\tilde{A} = 0.0072 \ \mu m^2/s^\alpha$ in the interval $\tilde{\varepsilon} \leq D_\alpha \leq \tilde{\kappa}$ with $\tilde{\varepsilon} = 0.005 \ \mu m^2/s^\alpha$ and $\tilde{\kappa} = 0.155 \ \mu m^2/s^\alpha$. Dimensionless numerical values of all quantities are used.

**Figure 2**

The normalized marginal distribution $g(\alpha)$. The histogram describes the probability to find a given value of the anomalous diffusion coefficient $\alpha$ and is based on the experimental data of Fig. 2(c) in Ref. [8]. The solid line shows the distribution of Eq. (2.5) with Eqs. (2.12) and (2.13): $a_0/c = 6.5 \ m^{-2}$ and $a_1 \beta_0/c = -2.03 \ m^{-2}$.

**Figure 3**

Log-log plot of the normalized distribution $\phi(D)$ of the diffusion constant $D$ in units of [$\mu\mathrm{m}^2/\mathrm{s}$]. The histogram is based on the experimental data in the left-bottom inset in Fig. 2(d) of Ref. [8]. The solid line shows the inverse gamma distribution of Eq. (B1) with $A = 0.011 \, \mu\mathrm{m}^2/\mathrm{s}$ in the interval $\varepsilon \leq D \leq \kappa$ with $\varepsilon = 0.025 \, \mu\mathrm{m}^2/\mathrm{s}$ and $\kappa = 0.205 \, \mu\mathrm{m}^2/\mathrm{s}$.

**Figure 4**

Semi-log plot of the normalized distribution $P(x,t)$ with variables $x\,[\mathrm{nm}]$ and $t\,[\mathrm{s}]$. This histogram uses the experimental data of Fig. 4(a) in Ref. [8]. The three lines show the $q$-Gaussian distributions of Eq. (2.9): $\hat{\alpha} = 0.53$ for the dashed line, $\hat{\alpha} = 0.57$ for the solid line, and $\hat{\alpha} = 0.59$ for the dotted line [8]. $|x| \leq 500\,\mathrm{nm}$ and $t = 9.2\,s$ ($\cong 0.41\,\mathrm{s}$) are considered.